\documentclass{aa}
\usepackage{epsfig}
\usepackage{txfonts}

\usepackage{epsfig}

\newcommand{\zsun}{\mbox{$Z_{\odot}$}}

\begin{document}

%\thesaurus{07(08.05.1; 08.13.2; 08.19.3; 08.23.3; 08.05.3)}

\title{In pursuit of gamma-ray burst progenitors: the identification of a sub-population of 
rotating Wolf-Rayet stars}

\author{Jorick S. Vink\inst{1}, G\"otz Gr\"afener\inst{1}, Tim J. Harries\inst{2}}
\offprints{Jorick S. Vink, jsv@arm.ac.uk}

 \institute{Armagh Observatory, College Hill, Armagh, BT61 9DG, Northern Ireland
            \and
School of Physics and Astronomy, University of Exeter, Stocker Rd, Exeter EX4 4QL, UK
}

\titlerunning{Rotating Wolf-Rayet stars}
\authorrunning{Jorick S. Vink}

\abstract{Long-duration gamma-ray bursts (GRBs) involve the most powerful cosmic explosions
since the Big Bang. Whilst it has been established that GRBs are 
related to the death throes of massive stars, the identification of 
their elusive progenitors has proved challenging. 
Theoretical modelling suggests that rotating Wolf-Rayet (WR) 
stars are the best candidates. 
Wolf-Rayet stars are thought to be in advanced 
core burning stages, just prior to explosion, but their strong stellar winds 
shroud their surfaces, preventing a direct measurement of their rotation. 
Fortunately, linear spectropolarimetry may be used to probe the flattening 
of their winds because of stellar spin. 
Spectropolarimetry surveys have shown that the vast majority of WR 
stars (80\%) have spherically symmetric winds and are therefore rotating slowly, yet 
a small minority (of 20\%) display a spectropolarimetric signature indicative 
of rotation. 
Here we find a highly significant correlation 
between WR objects that carry the signature of stellar rotation 
and the small subset of WR stars with ejecta nebulae that have 
only recently transitioned from a previous red sugergiant or luminous 
blue variable phase. 
As these youthful WR stars have yet to spin-down because of mass loss, they 
are the best candidate GRB progenitors identified to date. 
When we take recently published WR ejecta nebula numbers (of Stock \& Barlow 2010, MNRAS 409, 1429), 
we find that five out of the six line-effect WR stars are surrounded by ejecta nebulae. 
The statistics imply that the null hypothesis of no correlation between 
line-effect WR stars and ejecta nebulae can be rejected at the 
0.0004\% level. 
Given that four line-effect and WR ejecta nebula have 
spectroscopically been confirmed to contain nucleo-synthetic products, we argue 
that the correlation is both statistically significant and physically convincing.
The implication is that we have identified a sub-population of WR stars that 
fulfils the necessary criteria for making GRBs. 
Finally, we discuss the potential of identifying 
GRB progenitors via linear spectropolarimetry with extremely large telescopes.
}

\maketitle

%%%%%%%%%%%%%%%%%%%%%%%%%%%%%%%%%%%%%%%%%%%%%%%%%%%%%%%%%%%%%%%%%%%%%%%%%%%%%%%

\section{Introduction}
\label{s_intro}

Gamma-ray bursts (GRBs) comprise the most powerful cosmic explosions since the Big Bang. 
An overwhelming majority of GRBs are of the long and soft variety, which 
have been associated with stripped-envelope broad-lined supernovae SNe Ic 
(Galama et al. 1998; Hjorth et al. 2003; Stanek et al. 2003). 
GRBs represent the deaths of massive stars out to high redshifts, with 
the current record-holder lying at $z=9.4$ (Cucchiara et al. 2011).

Given the lack of hydrogen and helium in SN Ic spectra, 
the direct progenitors of SN Ic are widely expected to be stripped-envelope 
Wolf-Rayet (WR) stars with strong emission lines originating in 
dense stellar outflows. Even if some models involve binaries of lower mass stars 
to be the progenitors of H-poor Ibc SNe events, the broad-lined Ic are widely accepted 
to originate from WR stars (e.g. Modjaz et al. 2011). 
The most popular model for the production of a GRB involves the collapsar model 
where a rotating WR core collapses into a black hole.   
Yet, alternative GRB production models also invoke stellar rotation as the key 
ingredient for extracting the energy required to produce GRB jets (Woosley 2011).

Despite considerable progress in GRB research over the last decade, the actual 
identification of GRB progenitors has turned out to be a rather more challenging pursuit. 
Whilst it is generally expected that GRB progenitors are rotating WR stars, such 
objects have yet to be identified.  
The reason for this having been such a challenge is that it is generally not feasible to derive rotation 
rates from WR spectra. 
For main-sequence O-type stars it is possible to straightforwardly measure the projected 
equatorial rotational velocities ($v \sin i$) via the line-broadening effect of the 
stellar rotation on photospheric absorption lines, and 
O-type stars are indeed found to be mostly rotating rapidly (e.g. Howarth et al. 1997). 
Although angular momentum loss via stellar winds should slow this rapid rotation during massive 
star evolution, it is natural to assume that some WR stars could still rotate. 
Unfortunately, because of the nature of the strong winds from WR stars, all lines are in emission, and 
the traditional absorption-line method cannot be applied to WR stars. 

As an alternative to direct methods, one can search for deviations from sphericity of 
the WR stellar winds as a consequence of stellar rotation. 
These wind asymmetries would occur on spatial scales that are far too small to be resolved by 
direct imaging, or even the current generation of interferometers, but 
linear spectropolarimetry provides an elegant and unambiguous route 
to assessing wind asphericity. Continuum photons undergo electron 
scattering as they pass from the thermalization radius to the observer. 
If the projected scattering geometry is non-circular in the plane of the sky, the scattering 
introduces a small but measurable net linear polarization. 
The line photons, which are produced at larger radii than the continuum, see a smaller 
electron-scattering column and are thus less polarized than the continuum. 
The result is a depolarization at the emission line wavelengths. This effect has been observed 
in classical Be stars (e.g. Poeckert \& Marlborough 1976), O stars (Harries \& 
Howarth 1996; Vink et al. 2009), and WR stars.

\section{Spectropolarimetric surveys of WR stars}

More than a decade ago, Harries et al. (1998) 
investigated the incidence of line effects amongst WR stars. 
Out of a total sample of 29 Galactic WR stars, six were listed as line-effect objects. 
In other words, line-effect WR stars were found to be relatively rare, and 
it was suggested that the ``special'' property of the line-effect WR stars was related to 
stellar rotation. Harries et al. assumed that the photometric and spectroscopic variability
of the line-effect WR stars was related to the rotation period, which leads to surface rotation
rates of the order of 10\% of break-up ($\simeq$ 100\,km/s; Gr\"afener et al. in prep.)\footnote{We note that the stellar core might rotate significantly 
faster}, which may be fast enough to produce significant wind-compression effects according 
to the models of Ignace et al. (1996).

Close binary stars show intra-binary scattering that can lead to a measurable line-effect, 
for example CQ Cep (Harries \& Hilditch 1997), although the strength of the line effect 
changes as a function of orbital phase according to the second-harmonic of the period. Therefore the 
line effect may only be employed as an unambiguous wind diagnostic in WRs that are single or in binary 
systems where the separation is large enough that intra-binary scattering is negligible (see below).

Harries et al. surveyed the brightest northern hemisphere WR stars and found two 
single stars (WRs 134 and 137\footnote{WR137 is known to be a 
long (12 yr) period dust producing binary, but the binary separation rules out intra-binary 
scattering as the polarigenic source (Harries et al. 2000).}) that showed the 
so-called line-effect. The single star WR6 (EZ CMa) also has a strong, variable, line 
effect (Schulte-Ladbeck et al. 1991; Harries et al. 1999). 
Other observations of southern hemisphere WRs were presented by Schulte-Ladbeck 
(1994), who identified two additional single WR stars with a 
line-effect (WRs 16 and 40). Finally, the star WR136 was reported to have a line 
effect by Whitney et al. (1988).

\begin{figure*}
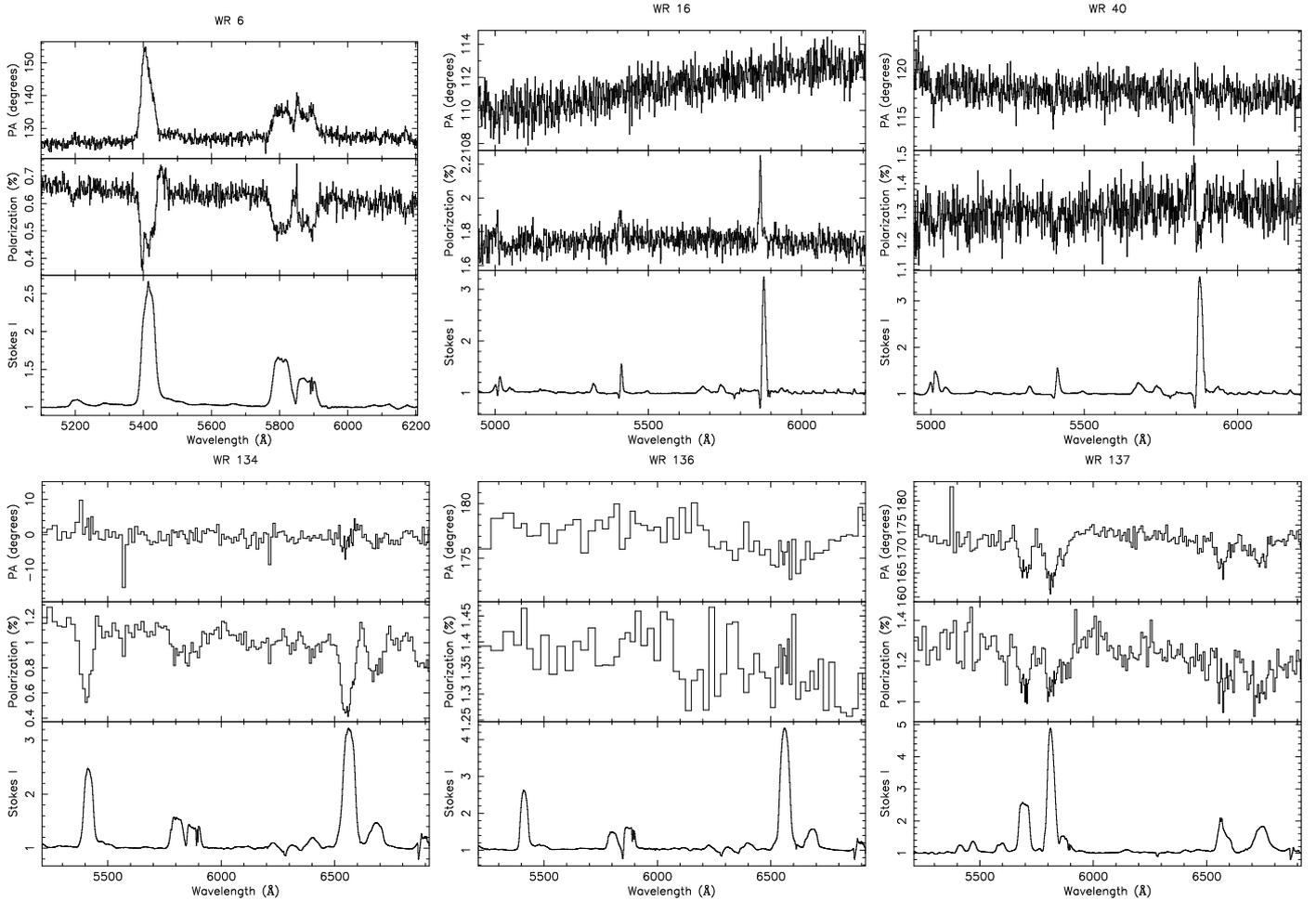

\begin{center}$
\begin{array}{ccc}
\includegraphics[width=6cm,clip]{wr6.ps} &
\includegraphics[width=6cm]{wr16.ps} &
\includegraphics[width=6cm]{wr40.ps} \\
\includegraphics[width=6cm]{wr134.ps} &
\includegraphics[width=6cm]{wr136.ps} &
\includegraphics[width=6cm]{wr137.ps}
\end{array}$
\end{center}
\caption{Galactic line-effect single WR stars (from Table 3 in Harries et al. 1998). 
The bottom section of each panel shows the normalized intensity spectrum, the middle 
sections show the polarization magnitude as a percentage, and the top sections 
show the polarization position angle in degrees. The polarization spectra have been 
binned to a constant error of 0.05\%. 
The data for WR 16 and WR 40 are from Schulte-Ladbeck (priv com), the remainder are 
from Harries et al. (1998).}
\label{line_effect_fig}
\end{figure*}

In Figure~\ref{line_effect_fig} we show all six single WR stars that have been 
reported to show a line-effect. WRs 6, 16 and 40 were observed using the RGO 
spectrograph on the 4-m Anglo-Australian telescope in March 1992. The other objects 
were observed using the ISIS spectrograph on the 4-m William Herschel Telescope during 
July 1993. The intensity spectra have been normalized using spline fits to interactively 
defined continuum regions, and the polarization spectra have been binned to a constant error of 0.05\%.

\begin{figure}
\begin{center}
\includegraphics[width=9cm,clip]{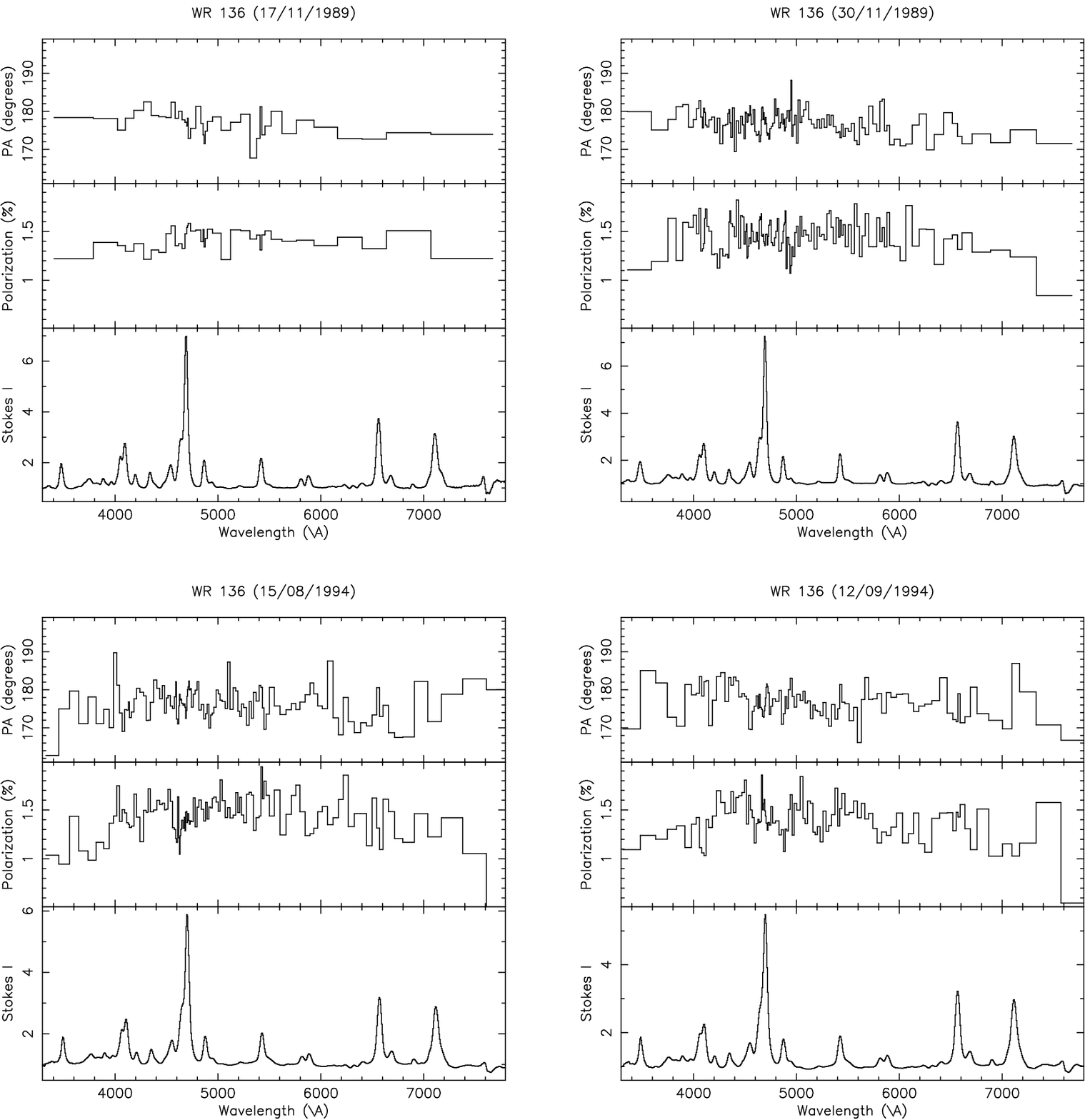} 
\end{center}
\caption{Spectropolarimetric data on WR 136 taken using the spectropolarimeter at Pine Bluff 
Observatory in 1989 and 1994. The plots are the same format as 
Figure~\ref{line_effect_fig} but the data are binned to a constant error of 0.1\%.}
\label{wr136_fig}
\end{figure}

Five of the six stars show a substantial depolarization, but WR136 does not. 
This was noted in Harries et al., but WR 136 was included in the line-effect stars, 
because a strong photo-polarimetric signature was shown in Whitney et al. (1988). 
For completeness, we have examined the archive 
spectropolarimetry from Pine Bluff Observatory (PBO). We 
find the evidence for a line effect in these noisy data to be marginal at best 
(see Figure~\ref{wr136_fig}). We conclude that WR 136 is a line-effect star with 
variable continuum polarization, and we urge follow-up
spectropolarimetry at high signal-to-noise, performed in a monitoring mode. 

Harries et al. (1998) were able to rule out the classical Be star scenario in which 
all objects are equally asymmetric, and with the observed polarization variation 
solely because of inclination effects. 
Although we cannot exclude the possibility that co-rotating interaction regions 
or other asymmetries could produce some of the 
linear polarization in certain cases, the most plausible 
scenario reproducing the observed linear polarization level {\it distribution} was 
shown to be one in which the intrinsic 
WR polarization levels were biased to low values. This implied that 
the vast majority (80\%) of WR stars have 
spherically symmetric winds indicative of slow rotation, whilst only a small (20\%) minority  
have significant intrinsic polarization (of $>$0.3\%). 
Radiative-transfer modelling demonstrated that these polarization levels were 
consistent with a wind asymmetry contrast between the low-
and high-density mass-loss regions of a factor of 2~to~3. 
Harries et al. concluded that the line-effect WR stars are the rotating ones.

A decade later, Vink (2007) considered the line-effect WR stars to constitute 
the most promising WR population for being GRB progenitors, and therefore studied an 
analogous sample in the lower metallicity environment of the Large Magellanic Cloud (LMC). 
Again, line-effect WR stars were found to be rare ($\sim$15\%), with, perhaps surprisingly, 
no apparent metallicity effect between the Galaxy and the LMC at an Fe content of roughly 50\% that of the 
Galactic value. 

\section{Ejecta nebulae around Wolf-Rayet stars}

Last year, Stock \& Barlow (2010) performed a search for ejecta nebulae around Galactic 
WR stars inspecting both the northern IPHAS H$\alpha$ survey (Drew et al. 2005) 
and the southern SHS H$\alpha$ survey (Parker et al. 2005). 
The fraction of Galactic WR stars with ejecta nebulae was claimed to be low ($\sim$6\%)\footnote{We argue that this number 
is a lower limit. According to our own criteria, the WR ejecta incidence is 23\% (see Gr\"afener et al. in prep), making the 
statistics more conservative, but still highly significant. 
We note that because the incidence is small, our conclusions are not sensitive to the precise criteria used to 
identify WR ejecta nebulae.}. 
A particularly interesting aspect concerning WR ejecta nebulae is that 
they contain nuclearly processed helium and/or nitrogen enriched material. 
Ejecta nebulae are thus thought to have formed from
strong mass-loss episodes during the prior red supergiant (RSG) or luminous blue variable (LBV)
phase, when the outer layers of the star might rotate slowly. 
If WR stars with ejecta nebulae have only recently transitioned from the RSG/LBV phase, 
one may possibly expect their central WR stars to be still rotating, before 
strong and persistent WR winds remove the remaining angular momentum, bringing them to an almost complete standstill. 

\section{A correlation between WR rotation and youth}

\begin{table}
\caption{The overlap between the WR ejecta nebula list of Stock \& Barlow (2010) and the 
line-effect stars according to Table 3 of Harries et al. (1998).}
\label{table}
\begin{tabular}{lll}
\hline
WR & Spec.Tp & Reported line-effect WR star? \\
\hline
6 & WN4 & yes\\
16& WN8 & yes\\
18& WN4 & no\\
40& WN8 & yes\\
134&WN6 & yes\\
136&WN6 & yes\\
\hline
\end{tabular}
\end{table}

One may ask the question whether the rare population of WR stars that possesses ejecta nebulae 
and the line-effect stars could be the same objects. 
The recent and independent list of Galactic WR ejecta nebulae is provided in 
Table 2 of Stock \& Barlow (2010). 
The Stock \& Barlow compilation and the Harries et al. (1998) 
study have just six objects in common, and as many as five of them show a line effect (see Table\,\ref{table}). 
Taking these numbers (of five out of six) at face value, a binomial test implies 
that the null hypothesis of no correlation between line-effect WR stars 
and ejecta nebulae can be rejected at the 0.000004 level (0.0004\%), 
i.e. the correlation is highly significant\footnote{If we were to apply our 
own Galactic WR ejecta nebula incidence rate from Gr\"afener et al. (in prep.), this number 
would increase but still be highly significant.}. 

Whilst one could possibly argue about morphological details of individual WR nebulae, such 
as whether a particular nebula does or does not constitute an unambiguous case for being an 
ejecta nebula (or merely a wind-blown one), given that four line-effect 
and Stock \& Barlow (2010) WR ejecta nebula have spectroscopically been confirmed to contain nucleo-synthetic products 
(see Table 1 in Stock \& Barlow; see also Stock et al. 2011), we argue 
that the correlation is both statistically significant and physically convincing. Finally, we 
refer to Gr\"afener et al. (in prep) for more detailed discussions, in particular regarding 
WR nebula morphologies and the relationship to circumstellar absorption features. 

\section{Summary and discussion}

We have found a statistically significant correlation 
between line-effect WR stars that have been postulated to be a rotating 
sub-population, and WR ejecta nebulae suggesting an 
evolutionary state just after the RSG/LBV phase. Our line-effect WR subset 
therefore represents the most promising group of objects fulfilling the criteria to 
potentially becoming a GRB.  

However, GRBs are rare events. Podsiadlowski et al. (2004) estimate 
that for every broad-lined SNIc-GRB there are at least another 100 non-GRB Type Ibc supernovae. 
This scarcity implies that another property (in addition to rotation rate 
and evolutionary state) is necessary for a WR star to produce a
GRB, for example low iron content (Vink \& de Koter 2005; Yoon \& Langer 2005; Woosley \& Heger 2006, Crowther 2006, 
Gr\"afener \& Hamann 2008).
This seems very plausible given the significant preference for long GRBs to occur in 
low-metallicity\footnote{Note however that these metallicity measurements involve  
oxygen rather than iron (Vink et al. 2001). For a discussion of the physics of 
Fe versus CNO line-driving see Vink et al. (1999) and Puls et al. (2000).} 
galaxies (e.g. Vreeswijk et al. 2004; Fruchter et al. 2006). This appears to be 
consistent with the tentative upper metallicity limit for GRB progenitors, as 
suggested by Vink (2007) on the basis of the low line-effect frequency in 
LMC WR stars, suggesting that the threshold metallicity, where significant 
differences in WR rotational properties occur, is below 0.5 $Z/\zsun$.

In order to discover the most likely population of long GRB progenitors, one would 
accordingly prefer to find low-metallicity rotating WR stars. 
Our results suggest that in principle this could be achieved either 
by the discovery of line-effect WR stars in low-metallicity galaxies, or by the 
search for WR ejecta nebulae in these systems.
Given the limited spatial resolution of traditional imaging 
studies, in practice spectropolarimetry on extremely large (30m class) telescopes 
appears to be the most promising avenue for directly pinpointing GRB 
progenitors -- provided these telescopes are equipped with the required polarization 
optics (Vink 2007, 2010). 

We have provided significant additional  
evidence that the line-effect WR stars are a rotating WR sub-group. 
This narrows down the group of WR stars that fulfil GRB progenitor criteria,  
making significant headway towards our final aim of directly pinpointing 
GRB progenitors. 

%*********************************************************************

\begin{acknowledgements}
We warmly thank Marilyn Meade from Pine Bluff Observatory and Regina Schulte-Ladbeck from 
the University of Pittsburgh for providing the spectropolarimetric data on WRs 16, 40, and 136.
\end{acknowledgements}


\begin{thebibliography}{}

\bibitem[]{}
Crowther P.A., 2006, ASPC 353, 157

\bibitem[]{}
Cucchiara A., Levan A.J., Fox D.B., et al., 2011, ApJ 736, 7

\bibitem[]{}
Drew J.E., Greimel R., Irwin M.J., et al., 2005, MNRAS 362, 753

\bibitem[]{}
Fruchter A.S., Levan A.J., Strolger L., et al., 2006, Nature 441, 463

\bibitem[]{}
Galama T.J., et al., 1998, Nature 395, 670

\bibitem[]{}
Gr\"afener G., \& Hamann W.-R. 2008, A\&A 482, 945

\bibitem[]{}
Harries T.J., \& Howarth I.D., 1996, A\&A 310, 533

\bibitem[]{}
Harries T.J., \& Hilditch R.W., 1997, MNRAS 291, 544

\bibitem[]{}
Harries T.J., Hillier D.J., \& Howarth I.D. 1998, MNRAS 296, 1072 

\bibitem[]{}
Harries T.J., Howarth I.D., Schulte-Ladbeck R.E., \& Hillier, D.J., 1999, MNRAS 302, 499
 
\bibitem[]{}
Harries T.J., Babler B.L., \& Fox G.K., 2000, A\&A 361, 273	

\bibitem[]{}
Hjorth J., Sollerman J., Moller P., et al., 2003, Nature 423, 847

\bibitem[]{}
Howarth I.D., Siebert K.W., Hussain G.A.J., \& Prinja R.K., 1997, MNRAS 284, 265

\bibitem[]{}
Ignace R., Cassinelli J.P., \& Bjorkman J.E., 1996, ApJ 459, 671

\bibitem[]{}
Modjaz M., Kewley L., Bloom J.S., et al., 2011, ApJ 731L, 4

\bibitem[]{}
Parker Q.A., Phillipps S., Pierce M.J., et al., 2005, MNRAS 362, 689

\bibitem[]{}
Podsiadlowski Ph., Mazzali P.A., Nomoto K., Lazzati D., \& Cappellaro E., 2004, ApJ 607L, 17

\bibitem[]{}
Poeckert R., \& Marlborough J.M., 1976, ApJ 206, 182

\bibitem[]{}
Puls J., Springmann U., \& Lennon M., 2000, A\&AS 141, 23

\bibitem[]{}
Schulte-Ladbeck R.E., et al., 1991, ApJ 382, 301

\bibitem[]{}
Schulte-Ladbeck R.E., 1994, Moffat A.F.J., Owocki S.P., Fullerton A.W., St-Louis N., eds., 
(Kluwer, Dordrecht), 347

\bibitem[]{}
Stanek K.Z., Matheson T.., Garnavich P.M., et al., 2003, ApJ 591, 17

\bibitem[]{}
Stock D.J., \& Barlow M.J., 2010, MNRAS 409, 1429

\bibitem[]{}
Stock D.J., Barlow M.J., 2011, \& Wesson R., astro-ph/1108.3800

\bibitem[]{}
Vink J.S., 2007, A\&A 469, 707

\bibitem[]{}
Vink J.S., 2010, Msngr 140 46

\bibitem[]{}
Vink J.S., \& de Koter, A. 2005, A\&A 442, 587

\bibitem[]{}
Vink J.S., de Koter A., \& Lamers H.J.G.L.M. 1999, A\&A 345, 109 

\bibitem[]{}
Vink J.S., de Koter A., \& Lamers H.J.G.L.M. 2001, A\&A 369, 574

\bibitem[]{}
Vink J.S., Davies B., Harries T.J., Oudmaijer R.D., \& Walborn N.R., 2009, A\&A 505, 743

\bibitem[]{}
Vreeswijk P.M., Ellison S.L., Ledoux C., et al. 2004, A\&A 419, 927 

\bibitem[]{}
Whitney B.A., et al., 1988, BAAS 20, 1013

\bibitem[]{}
Woosley S.E., 2011, astro-ph 1105.4193

\bibitem[]{}
Woosley S.E., \& Heger A. 2006, ApJ 637, 914 

\bibitem[]{}
Yoon S.-C., \& Langer N. 2005, A\&A, 443, 643 

\end{thebibliography}
\end{document}